\documentclass[twocolumn]{aastex631}

\shorttitle{T CrB as an extreme SU UMa type DN}
\shortauthors{I\l{}kiewicz et al.}
\graphicspath{{./}{figures/}}
\begin{document}

\title{Symbiotic star T CrB as an extreme SU UMa type dwarf nova}

\author{Krystian I\l{}kiewicz}
\affiliation{Astronomical Observatory, University of Warsaw, Al. Ujazdowskie 4, 00-478 Warszawa, Poland}

\author{Joanna Miko\l{}ajewska}
\affiliation{Nicolaus Copernicus Astronomical Center, Polish Academy of Sciences, Bartycka 18, 00716 Warsaw, Poland}

\author{Kiril A. Stoyanov}
\affiliation{Institute of Astronomy and National Astronomical Observatory, Bulgarian Academy of Sciences, Tsarigradsko Shose 72, BG-1784 Sofia, Bulgaria}

%% Note that the \and command from previous versions of AASTeX is now
%% depreciated in this version as it is no longer necessary. AASTeX 
%% automatically takes care of all commas and "and"s between authors names.

%% AASTeX 6.31 has the new \collaboration and \nocollaboration commands to
%% provide the collaboration status of a group of authors. These commands 
%% can be used either before or after the list of corresponding authors. The
%% argument for \collaboration is the collaboration identifier. Authors are
%% encouraged to surround collaboration identifiers with ()s. The 
%% \nocollaboration command takes no argument and exists to indicate that
%% the nearby authors are not part of surrounding collaborations.

%% Mark off the abstract in the ``abstract'' environment. 
\begin{abstract}

T~CrB is a symbiotic recurrent nova that exhibits quiescent and active phases between its classical nova eruptions. The statistical properties of these active phases have been poorly studied thus far. Because of that their nature remained unknown. Here we study statistical properties of the active phases and show that they are consistent with outburst and superoutbursts observed in SU~UMa type dwarf novae. The recurrence time of these outbursts is consistent with theoretical predictions for similar systems. Moreover, the visual and X-ray evolution of the last active phase is consistent with a superoutburst. This suggests that T~CrB is a dwarf nova with an extremely long orbital period, closely related to SU~UMa dwarf novae. The similarities between the last superoutburst and the reported activity preceding the 1946 nova eruption may suggest that next classical nova eruption in T~CrB could be indeed soon expected. 

\end{abstract}

%% Keywords should appear after the \end{abstract} command. 
%% The AAS Journals now uses Unified Astronomy Thesaurus concepts:
%% https://astrothesaurus.org
%% You will be asked to selected these concepts during the submission process
%% but this old "keyword" functionality is maintained in case authors want
%% to include these concepts in their preprints.
\keywords{Classical Novae --- Dwarf novae --- SU Ursae Majoris stars --- Symbiotic binary stars}

%% From the front matter, we move on to the body of the paper.
%% Sections are demarcated by \section and \subsection, respectively.
%% Observe the use of the LaTeX \label
%% command after the \subsection to give a symbolic KEY to the
%% subsection for cross-referencing in a \ref command.
%% You can use LaTeX's \ref and \label commands to keep track of
%% cross-references to sections, equations, tables, and figures.
%% That way, if you change the order of any elements, LaTeX will
%% automatically renumber them.
%%
%% We recommend that authors also use the natbib \citep
%% and \citet commands to identify citations.  The citations are
%% tied to the reference list via symbolic KEYs. The KEY corresponds
%% to the KEY in the \bibitem in the reference list below. 

\section{Introduction} \label{sec:intro}

T~CrB is a recurrent classical nova with a recurrence time of roughly 80 years \citep[e.g.][]{1949ApJ...109...81S}. The last recorded outburst was in 1946, which suggests a new outburst will occur in the next few years \citep[e.g.][]{2020ApJ...902L..14L}. Its orbital period is 227.5687d \citep{2000AJ....119.1375F}.     The white dwarf mass in T~CrB is close to the Chandrasekhar limit M$_{\mathrm{WD}}=1.2 \pm 0.2$~M$_\odot$ \citep{1998MNRAS.296...77B,2004A&A...415..609S}. The mass donor of the system is a red giant, classifying the system as a symbiotic star \citep[see][and references therein]{1986syst.book.....K}. This makes T~CrB a member of a rare group of symbiotic recurrent novae, which has only six members known thus far  \citep[RS~Oph, T~CrB, V3890~Sgr, V745~Sco, LMC~S154 and V618~Sgr; see][and references therein]{2019A&A...624A.133I,2023MNRAS.523..163M}. 

T~CrB experiences active phases, where the UV continuum of the system rises for extended periods of time \citep[e.g.][]{1992ApJ...393..289S,2016MNRAS.462.2695I}. The most recent active phase was suggested to be connected to the upcoming outburst \citep[e.g.][]{2016NewA...47....7M,2020ApJ...902L..14L,2023MNRAS.tmp..729S}. However, \citet{2016MNRAS.462.2695I} argued that the active phases are quasi-periodic and are not connected to the classical nova outbursts. Here we explore archival observations of T~CrB in order to unravel the nature of its active phases.

\section{Observations} \label{sec:obs}

\begin{figure*}%[h!]
\includegraphics[width=1.0\hsize]{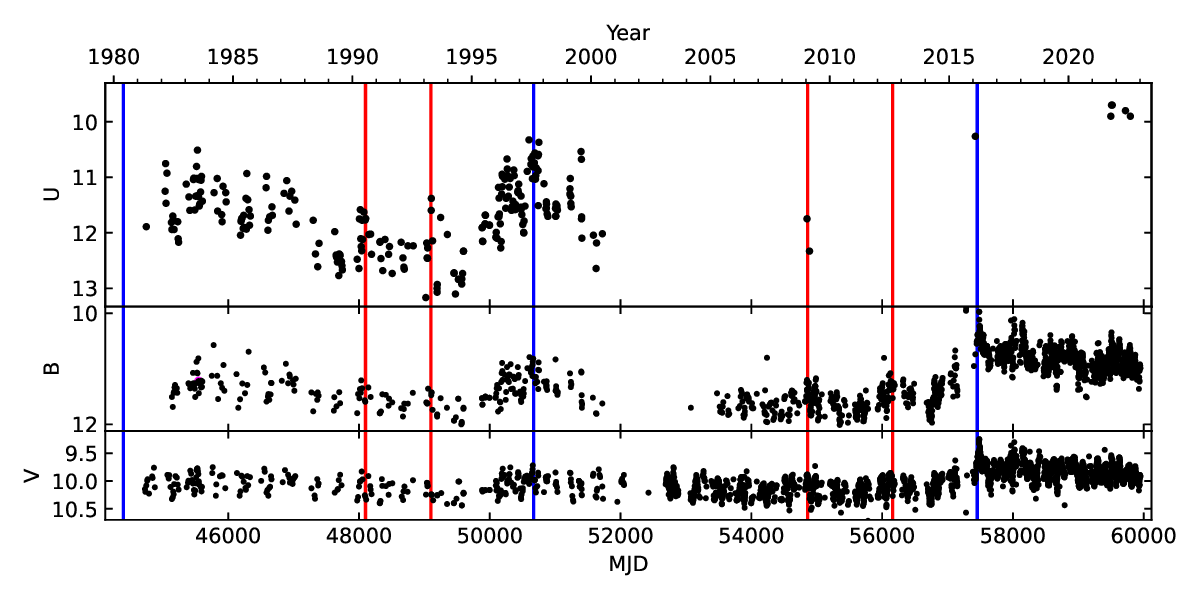}
\caption{The historical light-curve of T~CrB in $UBV$ filters. The blue lines indicate the start of superoutbursts and red lines indicate the times of normal dwarf nova outbursts (see text).}
\label{fig:raw}
\end{figure*}

We collected photometric observations in $UBV$ filters from the literature. The sources included the AAVSO International Database, the All Sky Automated Survey \citep{1997AcA....47..467P}, \citet{1992A&A...266..237B}, \citet{1997IBVS.4461....1Z}, \citet{2004MNRAS.350.1477Z}, \citet{2016NewA...47....7M} and \citet{2016ATel.8675....1Z}. In addition, we collected the data presented by \citet{2004A&A...415..609S} and references therein. Where not all data was in a machine-readable form we digitized the data in the same fashion as discussed by \citet{2004A&A...415..609S}. Our sample is virtually identical to that presented by \cite{2023MNRAS.tmp..729S}, except that we limited our analysis to times covered by observations in $U$, as the amplitude of active phases is best studied in bluest filters. The light-curve is presented in Fig.~\ref{fig:raw}.

\section{Results} \label{sec:results}

In order to examine the long-term variability of T~CrB we cleaned and smoothed the light-curve. As a first step we fitted the orbital variability in the $BV$ filters with a sine function using the data from before the start of the most recent active phase. From this we obtained the ephemeris of the maximum of the ellipsoidal variability MJD$_{\mathrm{max}}$=53155.22(49)+E$\times$113.71(2), implying an orbital period roughly consistent with the one measured by \citet{2000AJ....119.1375F}. The obtained amplitude of the ellipsoidal variability was 0.173(5)~mag in the $V$ band and 0.172(12) in the $B$ band. Subsequently, we removed the ellipsoidal variability, leaving only the mean red giant brightness in the data. We did not include the data in $U$ band in this procedure, as there was not enough data points for a good fit.

The resultant data was still noisy, as at all times in T~CrB there is flickering present with an amplitudes of up to 0.1~mag in the $B$ band \citep[e.g.][]{2016ATel.8675....1Z}. In order to remove such fast variability we smoothed the light-curve using local polynomial regression  \citep[LOESS,][]{cleveland1979robust,cleveland1988locally}. This method is commonly used to smooth astronomical light-curves \citep[e.g.][]{2012PASP..124..297L}. We used python implementation of LOESS, with a Gaussian kernel and a radius of 113.71~days \citep{sigvald_marholm_2019_3234461}. We note that \citet{2023MNRAS.tmp..729S} discovered that T~CrB orbital variability has slightly variable shape. This could affect the accuracy of our subtraction of the ellipsoidal variability from the light-curve. We tested this by changing the parameters of the subtracted orbital variability in accordance with the range of changes noted by  \citet{2023MNRAS.tmp..729S}. We found that the changing orbital variability does not significantly change the smoothed light-curve. 
%\footnote{\url{https://github.com/sigvaldm/localreg}}

\begin{figure*}%[h!]
\includegraphics[width=1.0\hsize]{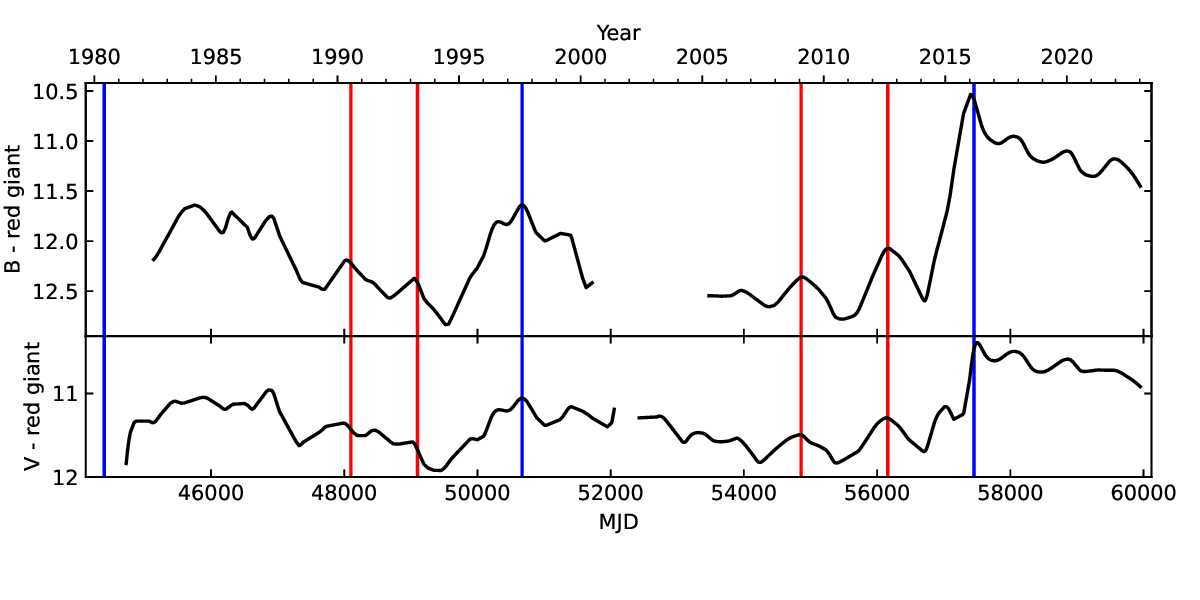}
\caption{The smoothed light-curve of T~CrB in $BV$ filters after the red giant contribution was removed. The blue lines indicate the start of superoutbursts and red lines indicate the times of normal dwarf nova outbursts (see text).}
\label{fig:smoothed}
\end{figure*}

In order to study the amplitude of long-term variability of T~CrB it is vital to remove the red giant contribution from the observed magnitudes, as its presence lowers the amplitude in the magnitude scale. In order to correct for that effect we adapted the mean brightness of the red giant of 10.029~mag in the $V$ band \citep{2004MNRAS.350.1477Z}. In order to calculate the red giant contribution in the $B$ band we assumed a reddening of E($B-V$)=0.15~mag \citep{1992ApJ...393..289S} and a color of the red giant $(B-V)_{\mathrm{RG}}$=1.60~mag \citep{1992msp..book.....S} assuming a M4.5III red giant \citep{1999A&AS..137..473M}. We then converted the theoretical magnitude of the red giant and observed magnitudes of T~CrB to fluxes, removed the red giant contribution from observations and converted the fluxes back to magnitudes. The resultant smoothed light-curve that has the red giant contribution removed is presented in Fig.~\ref{fig:smoothed}.

\citet{2016MNRAS.462.2695I} suggested that there are small and big active phases in T~CrB. The big active phases have a reccurence period of $\sim$5000~days. The small active phases occur between the big active phases, they have smaller amplitudes and a recurrence period of $\sim$1000~days. Both kinds of active phases have larger amplitudes in bluer filters. There are three big active phases in our data (Fig.~\ref{fig:smoothed}). However, the beginning of the first big active phase is not covered. Instead, in order to determine its maximum we used UV data presented by \citet{1992ApJ...393..289S}, which suggests that T~CrB reached a plateau around 8th June 1980. The maxima of the next two  big active phases were determined from our smoothed light-curve to be around MJD 50670 and 57455. The first two maxima of small active phases were determined to be around MJD 48100 and 49100 by \citet{1997IBVS.4461....1Z}. The maxima of the two most recent small active phases were determined from our smoothed light-curve to be around MJD 54860 and 56160. We measured the amplitudes of the big and small active phases using the smoothed light-curve in the $B$ band. The measurements are presented in Table~\ref{tab:active}.

\begin{table}[h]
\centering
\caption{The time of maxima and amplitudes in the $B$ band of big and small active phases in T~CrB.} \label{tab:active}
\begin{tabular}{ccc}
\tablewidth{0pt}
\hline
\hline
Type  &		MJD &	Amplitude		\\
\hline
\decimals

Big active phase / superoutburst & 44398 & 1.17 mag\\ %12.77-11.60
Small active phase / normal outburst & 48100 & 0.63 mag\\ %12.77-12.14
Small active phase / normal outburst & 49100 & 0.47 mag\\ %12.77-12.30
Big active phase / superoutburst & 50670 & 1.15 mag\\ %12.73-11.58
Small active phase / normal outburst & 54860 & 0.43 mag\\ %12.73-12.3
Small active phase / normal outburst & 56160 & 0.73 mag\\ %12.73-12.00
Big active phase / superoutburst & 57455 & 1.99 mag\\ %12.60-10.61
\hline
\end{tabular}
\end{table}

%Asqw= 12.60-11.03

Dwarf nova outbursts as a source of active phases in T~CrB have been suggested in the past \citep{2016MNRAS.462.2695I}. Moreover, the X-ray flux dropped right at the beginning of the most recent big active phase and returned to normal right at the end of the active phase \citep{2018A&A...619A..61L,2023ATel16114....1K}. This is exactly the behavior expected for a dwarf nova outburst \citep[e.g][]{1645d388-dd61-39f9-b8ad-83955da1c695,2018A&A...619A..61L}. Exactly the same behavior was observed during the previous big active phase, were the X-ray flux increased at the end of the big active phase \citep{2016MNRAS.462.2695I}. This confirms that both active phases are of the same nature and are reminiscent of a dwarf nova outburst.

The photometric evolution of the most recent big active phase evolution is very similar to a typical superoutburst. Namely, fast rise to the maximum is followed a quasi-plateau with a slow decline, and subsequent fast drop to quiescence (Fig.~\ref{fig:smoothed}, see also fig.~1 of \citealt{2016MNRAS.460.2526O}). While the sharp decline to quiescence is not yet fully covered, it has already started \citep{2023ATel16107....1S,2023ATel16109....1T,2023arXiv230700255M}.

\begin{figure*}
\centering
\includegraphics[width=0.49\hsize]{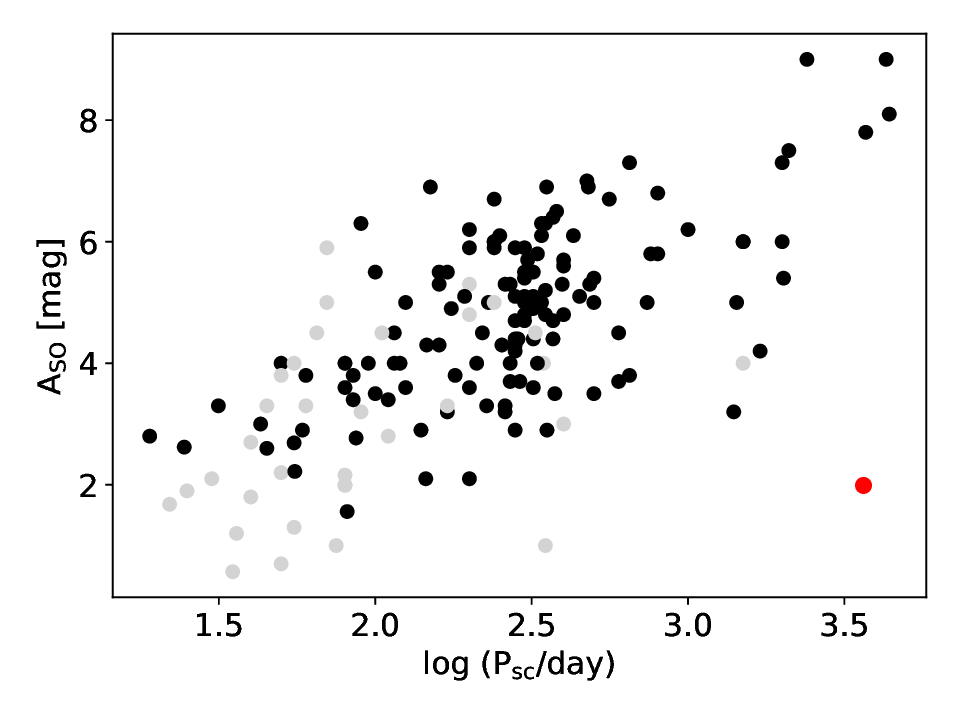}
\includegraphics[width=0.49\hsize]{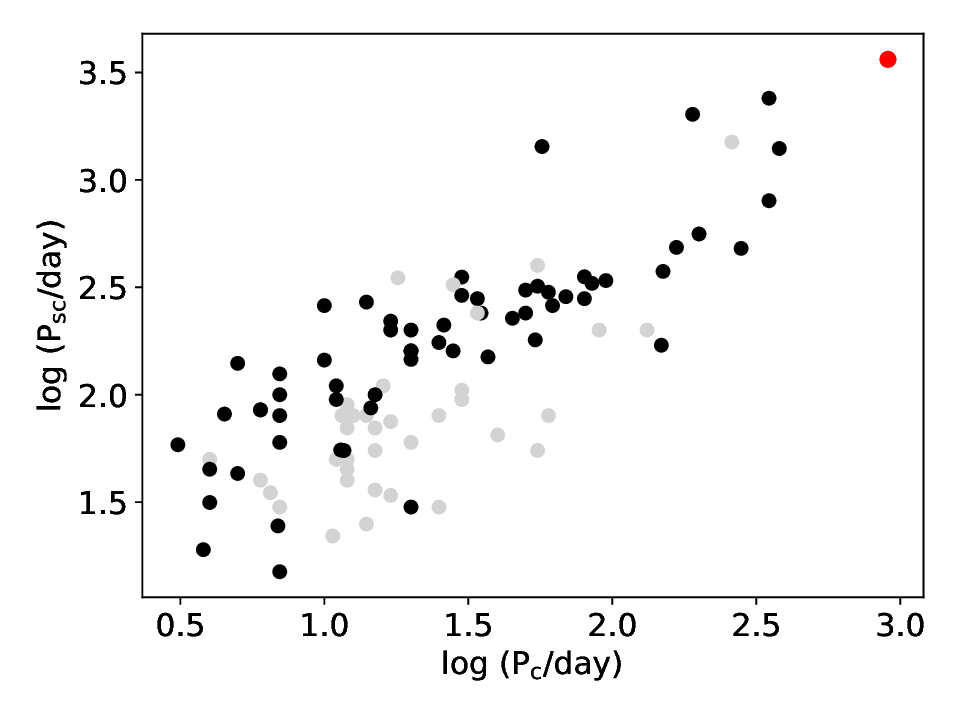}
\includegraphics[width=0.49\hsize]{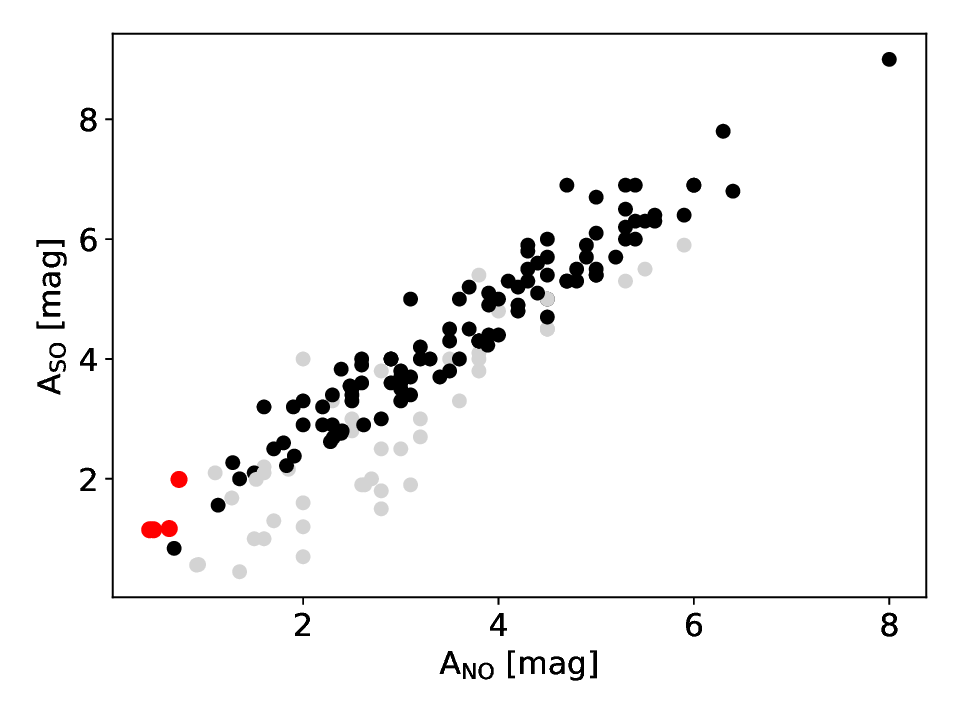}
\includegraphics[width=0.49\hsize]{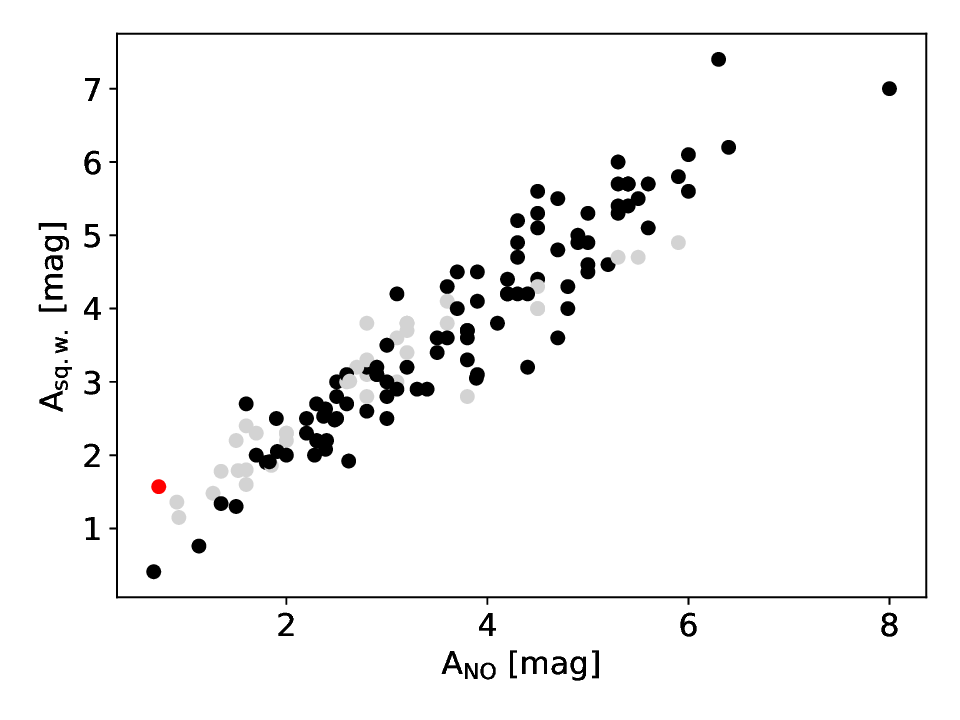}

\caption{Relationships between normal dwarf nova/short dwarf nova outbursts and superoutbursts/long dwarf nova outbursts properties \citep{2016MNRAS.460.2526O}. The SU~UMa type dwarf novae are plotted with the black points, the remaining dwarf novae are plotted with the gray points and the measurements for T~CrB are plotted with the red points (see text). }
 \label{fig:otulakowska}
\end{figure*}

In order to test the hypothesis that active phases in T~CrB are due to dwarf nova outbursts  we compared their statistical properties with a sample of dwarf novae. In this hypothesis the small active phases are the normal dwarf nova outbursts and big active phases are superoubtursts in SU~UMa type dwarf novae. In SU~UMa type dwarf novae the superoutburst occurs because of a 3:1 tidal resonance in the accretion disc \citep{1990PASJ...42..135H,2005PJAB...81..291O}. The statistical properties of dwarf novae selected for the present test consisted of relationships between observables that had the highest correlation among those presented by  \citet{2016MNRAS.460.2526O}. In particular, we compared amplitudes of normal outbursts A$_{\mathrm{NO}}$ and superoutbursts A$_{\mathrm{SO}}$, $minimal$ recurrence time of normal outbursts P$_{\mathrm{c}}$ and super outburst P$_{\mathrm{sc}}$, as well as amplitude of the superoutburst when magnitude from the end of the plateau is used A$_{\mathrm{sq.w.}}$. Both P$_{\mathrm{c}}$ and P$_{\mathrm{sc}}$ are not constant and can change by a factor of up to 3 \citep{2016MNRAS.460.2526O}. Hence, it is essential to use minimal recurrence time observed. For this reason we do not adapt recurrence times measured by us, but minimal values recorded in the literature. Namely, we use P$_{\mathrm{c}}$=906d and  P$_{\mathrm{sc}}$=3640d measured by \citet{1997ppsb.conf..117A}.  For amplitudes we adopted the values presented in Table~\ref{tab:active}. As for A$_{\mathrm{sq.w.}}$ it is possible only to measure it during the last big active phase and we obtained A$_{\mathrm{sq.w.}}$=1.57~mag.

The comparison of active phases in T~CrB to those in dwarf novae is presented in Fig.~\ref{fig:otulakowska}. On all of the figures the T~CrB active phases are consistent with the dwarf nova properties. This, combined with the X-ray behavior and the fact that the active phases behavior remained roughly constant for at least the last $\sim$40~years confirms their dwarf nova nature. The only relationship that T~CrB slightly departs from is the relationship between P$_{\mathrm{sc}}$ and A$_{\mathrm{SO}}$. This may be explained by a small sample of measured recurrence times of big active phases. Moreover, this relationship has the lowest Pearson product-moment correlation coefficient of r=0.65 compared to r=0.84, r=0.94 and r=0.89 for other relationships.

\begin{figure}[h!]
\centering
\includegraphics[width=0.99\hsize]{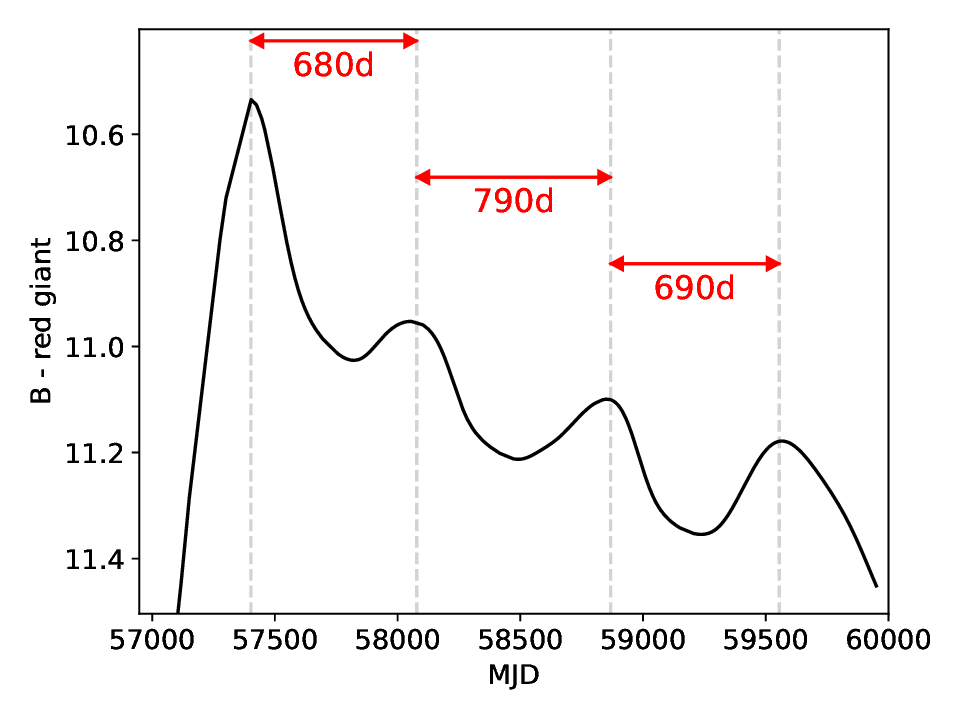}

\caption{Same as Fig.~\ref{fig:smoothed}, but zoomed in on the most recent superoutburst. The maxima of suspected superhumps are marked with dashed lines.}
 \label{fig:superhumps}
\end{figure}

During superoutbursts in SU~UMa dwarf novae superhumps are observed. Superhumps have roughly sinusoidal shape and periods slightly larger compared to the orbital period \citep[e.g.][]{2009PASJ...61S.395K}. In fact, during the plateau of the last big active phase of T~CrB a variability with a period of $\sim$700d is visible (Fig.~\ref{fig:superhumps}). A similar behaviour might have been present during the two previous superoutbursts, but not enough data is available to confirm this. Since the orbital period of T~CrB is 227.5687d \citep{2000AJ....119.1375F} the $\sim$700d is significantly larger than the orbital period and we suggest that they can be interpreted as superhumps. This would further confirm the  SU~UMa type dwarf nova nature of the active phases in T~CrB. We note that this period excess of the positive superhump is unprecedentedly high, being over three times the orbital period. Typically a period excess of up to 10\% is observed \citep[e.g.][]{2021PASJ...73.1209W}. This is easily explained by the fact that the period excess of a superhump is strongly correlated with the orbital period \citep{1984A&A...132..187S} and the dwarf novae were observed in binary systems with orbital periods of up to few hours \citep[e.g.][]{2023arXiv230413311K}. In fact, if we would extrapolate the known relationship between the period excess and orbital period, we would obtain period excees far exceeding the one observed in T~CrB \citep[see eq. 4 of][]{2016MNRAS.460.2526O}. However, obviously an extrapolation so large will not provide an accurate prediction. Moreover, the rationship between orbital period and superhump period, as well as a similar relation between superhump period and mass ratio in the system seem to not hold for systems with long orbital periods \citep[see][and references therein]{2016MNRAS.460.2526O}. Hence, the validity of superhump interpretation of variability with $\sim$700d period should be verified with theoretical simulations.

~\\

\section{Conclusions and discussion} \label{sec:concl}

It has been suggested that the last T~CrB big active phase is biggest since the 1946 outburst and the last similar big active phase was seen just before the 1946 outbursts \citep{2016NewA...47....7M,2020ApJ...902L..14L,2023MNRAS.tmp..729S}. This may be true regardless of physical interpretation of the brightening and classical nova eruption could be triggered by a superoutburst. Moreover, we note that a systematic change of mass transfer rate in the system suggested by \citet{2023MNRAS.tmp..729S} is consistent with a change in recurrence time of superoutbursts from 3640d  \citep{1997ppsb.conf..117A}  to $\sim$6500d (Table~\ref{tab:active}). However, the change in the amplitude of superoutbursts is not as obvious given that most of the historical data is in the $V$ band, where the amplitude of outburst is relatively low.  On the other hand, the limited sample of $U$ observations suggests that the most recent superoutburst had an amplitude only $\sim$0.5~mag higher compared to the previous two superoutbursts (Fig.~\ref{fig:raw}). This makes the difference in amplitudes less dramatic compared to what is suggested in the literature. Moreover, Variable Star and Exoplanet Section of Czech Astronomical Society\footnote{\url{http://var2.astro.cz/}} provides observations of T~CrB suggesting that the most recent superoutburst had nearly identical maximum brightness as the preceding one. However, we omitted these observations in our work as they seem inconsistent with the AAVSO data.

We have shown that T~CrB is closely related to SU~UMa dwarf novae using statistical properties of its active phases, the shape of the most recent big active phase, as well as the X-ray behavior. This makes T~CrB a dwarf nova with an extremely long orbital period, $\sim$3 orders of magnitude longer than in other SU~UMa type dwarf novae. This also places T~CrB among rare systems where both dwarf nova and classical nova outbursts were recorded \citep{2002A&A...382..910S,2016Natur.537..649M}. 

Dwarf nova outbursts in symbiotic stars have been modeled in the past \citep[e.g.][]{2011MNRAS.418.2576A}.  In particular, according to the recent modeling by \citet{2018MNRAS.481.5422B} the recurrence time of dwarf nova outbursts in similar symbiotic stars is expected to be in the range between 1.5 and 4~years. This is in excellent agreement with the $\sim$1000d (2.7~years) recurrence time of normal dwarf nova outbursts in T~CrB. Hence, the dwarf nova outbursts in T~CrB are consistent with theoretical predictions. This makes T~CrB a valuable laboratory for dwarf nova outburst models in extremely large accretion discs. Moreover, it may serve as a link between dwarf nova outbursts in binary stars and similar outbursts in Active Galactic Nuclei \citep{2009A&A...496..413H}.

\begin{acknowledgments}
We acknowledge with thanks the variable star observations from the AAVSO International Database contributed by observers worldwide and used in this research. This work was supported by Polish National Science Center grants Sonatina 2021/40/C/ST9/00186 and OPUS 2017/27/B/01940.
\end{acknowledgments}

\bibliography{reference}{}

\begin{thebibliography}{}
\expandafter\ifx\csname natexlab\endcsname\relax\def\natexlab#1{#1}\fi
\providecommand{\url}[1]{\href{#1}{#1}}
\providecommand{\dodoi}[1]{doi:~\href{http://doi.org/#1}{\nolinkurl{#1}}}
\providecommand{\doeprint}[1]{\href{http://ascl.net/#1}{\nolinkurl{http://ascl.net/#1}}}
\providecommand{\doarXiv}[1]{\href{https://arxiv.org/abs/#1}{\nolinkurl{https://arxiv.org/abs/#1}}}

\bibitem[{{Alexander} {et~al.}(2011){Alexander}, {Wynn}, {King}, \&
  {Pringle}}]{2011MNRAS.418.2576A}
{Alexander}, R.~D., {Wynn}, G.~A., {King}, A.~R., \& {Pringle}, J.~E. 2011,
  \mnras, 418, 2576, \dodoi{10.1111/j.1365-2966.2011.19647.x}

\bibitem[{{Anupama}(1997)}]{1997ppsb.conf..117A}
{Anupama}, G.~C. 1997, in Physical Processes in Symbiotic Binaries and Related
  Systems, ed. J.~{Miko{\l}ajewska}, 117

\bibitem[{{Belczynski} \& {Mikolajewska}(1998)}]{1998MNRAS.296...77B}
{Belczynski}, K., \& {Mikolajewska}, J. 1998, \mnras, 296, 77,
  \dodoi{10.1046/j.1365-8711.1998.01301.x}

\bibitem[{{Bollimpalli} {et~al.}(2018){Bollimpalli}, {Hameury}, \&
  {Lasota}}]{2018MNRAS.481.5422B}
{Bollimpalli}, D.~A., {Hameury}, J.~M., \& {Lasota}, J.~P. 2018, \mnras, 481,
  5422, \dodoi{10.1093/mnras/sty2555}

\bibitem[{{Bruch}(1992)}]{1992A&A...266..237B}
{Bruch}, A. 1992, \aap, 266, 237

\bibitem[{Cleveland(1979)}]{cleveland1979robust}
Cleveland, W.~S. 1979, Journal of the American statistical association, 74, 829

\bibitem[{Cleveland \& Devlin(1988)}]{cleveland1988locally}
Cleveland, W.~S., \& Devlin, S.~J. 1988, Journal of the American statistical
  association, 83, 596

\bibitem[{{Fekel} {et~al.}(2000){Fekel}, {Joyce}, {Hinkle}, \&
  {Skrutskie}}]{2000AJ....119.1375F}
{Fekel}, F.~C., {Joyce}, R.~R., {Hinkle}, K.~H., \& {Skrutskie}, M.~F. 2000,
  \aj, 119, 1375, \dodoi{10.1086/301260}

\bibitem[{Fertig {et~al.}(2011)Fertig, Mukai, Nelson, \&
  Cannizzo}]{1645d388-dd61-39f9-b8ad-83955da1c695}
Fertig, D., Mukai, K., Nelson, T., \& Cannizzo, J.~K. 2011, Publications of the
  Astronomical Society of the Pacific, 123, 1054.
\newblock \url{http://www.jstor.org/stable/10.1086/661949}

\bibitem[{{Hameury} {et~al.}(2009){Hameury}, {Viallet}, \&
  {Lasota}}]{2009A&A...496..413H}
{Hameury}, J.~M., {Viallet}, M., \& {Lasota}, J.~P. 2009, \aap, 496, 413,
  \dodoi{10.1051/0004-6361/200810928}

\bibitem[{{Hirose} \& {Osaki}(1990)}]{1990PASJ...42..135H}
{Hirose}, M., \& {Osaki}, Y. 1990, \pasj, 42, 135

\bibitem[{{I{\l}kiewicz} {et~al.}(2019){I{\l}kiewicz}, {Miko{\l}ajewska},
  {Miszalski}, {Gromadzki}, {Monard}, \& {Amigo}}]{2019A&A...624A.133I}
{I{\l}kiewicz}, K., {Miko{\l}ajewska}, J., {Miszalski}, B., {et~al.} 2019,
  \aap, 624, A133, \dodoi{10.1051/0004-6361/201834165}

\bibitem[{{I{\l}kiewicz} {et~al.}(2016){I{\l}kiewicz}, {Miko{\l}ajewska},
  {Stoyanov}, {Manousakis}, \& {Miszalski}}]{2016MNRAS.462.2695I}
{I{\l}kiewicz}, K., {Miko{\l}ajewska}, J., {Stoyanov}, K., {Manousakis}, A., \&
  {Miszalski}, B. 2016, \mnras, 462, 2695, \dodoi{10.1093/mnras/stw1837}

\bibitem[{{Kato} \& {Vanmunster}(2023)}]{2023arXiv230413311K}
{Kato}, T., \& {Vanmunster}, T. 2023, arXiv e-prints, arXiv:2304.13311,
  \dodoi{10.48550/arXiv.2304.13311}

\bibitem[{{Kato} {et~al.}(2009){Kato}, {Imada}, {Uemura}, {Nogami}, {Maehara},
  {Ishioka}, {Baba}, {Matsumoto}, {Iwamatsu}, {Kubota}, {Sugiyasu}, {Soejima},
  {Moritani}, {Ohshima}, {Ohashi}, {Tanaka}, {Sasada}, {Arai}, {Nakajima},
  {Kiyota}, {Tanabe}, {Imamura}, {Kunitomi}, {Kunihiro}, {Taguchi}, {Koizumi},
  {Yamada}, {Nishi}, {Kida}, {Tanaka}, {Ueoka}, {Yasui}, {Maruoka}, {Henden},
  {Oksanen}, {Moilanen}, {Tikkanen}, {Aho}, {Monard}, {Itoh}, {Dubovsky},
  {Kudzej}, {Dancikova}, {Vanmunster}, {Pietz}, {Bolt}, {Boyd}, {Nelson},
  {Krajci}, {Cook}, {Torii}, {Starkey}, {Shears}, {Jensen}, {Masi}, {Hynek},
  {Nov{\'a}k}, {Koci{\'a}n}, {Kr{\'a}l}, {Ku{\v{c}}{\'a}kov{\'a}}, {Kolasa},
  {{\v{S}}tastn{\'y}}, {Staels}, {Miller}, {Sano}, {Ponthi{\`e}re},
  {Miyashita}, {Crawford}, {Brady}, {Santallo}, {Richards}, {Martin},
  {Buczynski}, {Richmond}, {Kern}, {Davis}, {Crabtree}, {Beaulieu}, {Davis},
  {Aggleton}, {Morelle}, {Pavlenko}, {Andreev}, {Baklanov}, {Koppelman},
  {Billings}, {Urban{\v{c}}ok}, {{\"O}gmen}, {Heathcote}, {Gomez}, {Voloshina},
  {Retter}, {Mularczyk}, {Z{\l}oczewski}, {Olech}, {Kedzierski}, {Pickard},
  {Stockdale}, {Virtanen}, {Morikawa}, {Hambsch}, {Garradd}, {Gualdoni},
  {Geary}, {Omodaka}, {Sakai}, {Michel}, {C{\'a}rdenas}, {Gazeas}, {Niarchos},
  {Yushchenko}, {Mallia}, {Fiaschi}, {Good}, {Walker}, {James}, {Douzu},
  {Julian}, {Butterworth}, {Shugarov}, {Volkov}, {Chochol}, {Katysheva},
  {Rosenbush}, {Khramtsova}, {Kehusmaa}, {Reszelski}, {Bedient}, {Liller},
  {Pojma{\'n}ski}, {Simonsen}, {Stubbings}, {Schmeer}, {Muyllaert}, {Kinnunen},
  {Poyner}, {Ripero}, \& {Kriebel}}]{2009PASJ...61S.395K}
{Kato}, T., {Imada}, A., {Uemura}, M., {et~al.} 2009, \pasj, 61, S395,
  \dodoi{10.1093/pasj/61.sp2.S395}

\bibitem[{{Kenyon}(1986)}]{1986syst.book.....K}
{Kenyon}, S.~J. 1986, {The symbiotic stars}

\bibitem[{{Kuin} {et~al.}(2023){Kuin}, {Luna}, {Page}, {Mukai}, {Sokoloski},
  {Osborne}, \& {Schaefer}}]{2023ATel16114....1K}
{Kuin}, N.~P., {Luna}, G. J.~M., {Page}, K., {et~al.} 2023, The Astronomer's
  Telegram, 16114, 1

\bibitem[{{Li} {et~al.}(2012){Li}, {Chen}, \& {Wang}}]{2012PASP..124..297L}
{Li}, Z., {Chen}, L., \& {Wang}, D. 2012, \pasp, 124, 297,
  \dodoi{10.1086/665327}

\bibitem[{{Luna} {et~al.}(2020){Luna}, {Sokoloski}, {Mukai}, \& {M.
  Kuin}}]{2020ApJ...902L..14L}
{Luna}, G. J.~M., {Sokoloski}, J.~L., {Mukai}, K., \& {M. Kuin}, N.~P. 2020,
  \apjl, 902, L14, \dodoi{10.3847/2041-8213/abbb2c}

\bibitem[{{Luna} {et~al.}(2018){Luna}, {Mukai}, {Sokoloski}, {Nelson}, {Kuin},
  {Segreto}, {Cusumano}, {Jaque Arancibia}, \&
  {Nu{\~n}ez}}]{2018A&A...619A..61L}
{Luna}, G.~J.~M., {Mukai}, K., {Sokoloski}, J.~L., {et~al.} 2018, \aap, 619,
  A61, \dodoi{10.1051/0004-6361/201833747}

\bibitem[{Marholm(2019)}]{sigvald_marholm_2019_3234461}
Marholm, S. 2019, {sigvaldm/localreg: Fully tested version}, 0.2.0,  Zenodo,
  \dodoi{10.5281/zenodo.3234461}

\bibitem[{{Merc} {et~al.}(2023){Merc}, {G{\'a}lis}, {Velez}, {Charbonnel},
  {Garde}, {Le D{\^u}}, {Mulato}, {Petit}, {Bohlsen}, {Curry}, {Love}, \&
  {Barker}}]{2023MNRAS.523..163M}
{Merc}, J., {G{\'a}lis}, R., {Velez}, P., {et~al.} 2023, \mnras, 523, 163,
  \dodoi{10.1093/mnras/stad1434}

\bibitem[{{Mr{\'o}z} {et~al.}(2016){Mr{\'o}z}, {Udalski}, {Pietrukowicz},
  {Szyma{\'n}ski}, {Soszy{\'n}ski}, {Wyrzykowski}, {Poleski}, {Koz{\l}owski},
  {Skowron}, {Ulaczyk}, {Skowron}, \& {Pawlak}}]{2016Natur.537..649M}
{Mr{\'o}z}, P., {Udalski}, A., {Pietrukowicz}, P., {et~al.} 2016, \nat, 537,
  649, \dodoi{10.1038/nature19066}

\bibitem[{{Munari}(2023)}]{2023arXiv230700255M}
{Munari}, U. 2023, arXiv e-prints, arXiv:2307.00255,
  \dodoi{10.48550/arXiv.2307.00255}

\bibitem[{{Munari} {et~al.}(2016){Munari}, {Dallaporta}, \&
  {Cherini}}]{2016NewA...47....7M}
{Munari}, U., {Dallaporta}, S., \& {Cherini}, G. 2016, \na, 47, 7,
  \dodoi{10.1016/j.newast.2016.01.002}

\bibitem[{{M{\"u}rset} \& {Schmid}(1999)}]{1999A&AS..137..473M}
{M{\"u}rset}, U., \& {Schmid}, H.~M. 1999, \aaps, 137, 473,
  \dodoi{10.1051/aas:1999105}

\bibitem[{{Osaki}(2005)}]{2005PJAB...81..291O}
{Osaki}, Y. 2005, Proceedings of the Japan Academy, Series B, 81, 291,
  \dodoi{10.2183/pjab.81.291}

\bibitem[{{Otulakowska-Hypka} {et~al.}(2016){Otulakowska-Hypka}, {Olech}, \&
  {Patterson}}]{2016MNRAS.460.2526O}
{Otulakowska-Hypka}, M., {Olech}, A., \& {Patterson}, J. 2016, \mnras, 460,
  2526, \dodoi{10.1093/mnras/stw1120}

\bibitem[{{Pojmanski}(1997)}]{1997AcA....47..467P}
{Pojmanski}, G. 1997, \actaa, 47, 467, \dodoi{10.48550/arXiv.astro-ph/9712146}

\bibitem[{{Sanford}(1949)}]{1949ApJ...109...81S}
{Sanford}, R.~F. 1949, \apj, 109, 81, \dodoi{10.1086/145106}

\bibitem[{{Schaefer}(2023)}]{2023MNRAS.tmp..729S}
{Schaefer}, B.~E. 2023, \mnras, \dodoi{10.1093/mnras/stad735}

\bibitem[{{Schaefer} {et~al.}(2023){Schaefer}, {Kloppenborg}, {Waagen}, \&
  {Observers}}]{2023ATel16107....1S}
{Schaefer}, B.~E., {Kloppenborg}, B., {Waagen}, E.~O., \& {Observers}, T.~A.
  2023, The Astronomer's Telegram, 16107, 1

\bibitem[{{Selvelli} {et~al.}(1992){Selvelli}, {Cassatella}, \&
  {Gilmozzi}}]{1992ApJ...393..289S}
{Selvelli}, P.~L., {Cassatella}, A., \& {Gilmozzi}, R. 1992, \apj, 393, 289,
  \dodoi{10.1086/171506}

\bibitem[{{Stanishev} {et~al.}(2004){Stanishev}, {Zamanov}, {Tomov}, \&
  {Marziani}}]{2004A&A...415..609S}
{Stanishev}, V., {Zamanov}, R., {Tomov}, N., \& {Marziani}, P. 2004, \aap, 415,
  609, \dodoi{10.1051/0004-6361:20034623}

\bibitem[{{Stolz} \& {Schoembs}(1984)}]{1984A&A...132..187S}
{Stolz}, B., \& {Schoembs}, R. 1984, \aap, 132, 187

\bibitem[{{Strai{\v{z}}ys}(1992)}]{1992msp..book.....S}
{Strai{\v{z}}ys}, V. 1992, {Multicolor stellar photometry}

\bibitem[{{Teyssier} {et~al.}(2023){Teyssier}, {Hinnefeld}, {Boussin},
  {Diabassoura}, {Guarro Flo}, {Sims}, {Leduc}, {Curry}, {Boyd}, {Cujedo}, \&
  {Shore}}]{2023ATel16109....1T}
{Teyssier}, F., {Hinnefeld}, J.~D., {Boussin}, C., {et~al.} 2023, The
  Astronomer's Telegram, 16109, 1

\bibitem[{{{\v{S}}imon}(2002)}]{2002A&A...382..910S}
{{\v{S}}imon}, V. 2002, \aap, 382, 910, \dodoi{10.1051/0004-6361:20011560}

\bibitem[{{Wakamatsu} {et~al.}(2021){Wakamatsu}, {Thorstensen}, {Kojiguchi},
  {Isogai}, {Kimura}, {Ohnishi}, {Kato}, {Itoh}, {Sugiura}, {Sumiya},
  {Matsumoto}, {Ito}, {Nikai}, {Akitaya}, {Ishioka}, {Oide}, {Kanai}, {Uzawa},
  {Oasa}, {Tordai}, {Vanmunster}, {Shugarov}, {Yamanaka}, {Sasada}, {Takagi},
  {Nishinaka}, {Yamazaki}, {Otsubo}, {Nakaoka}, {Murata}, {Ohsawa}, {Morita},
  {Ichiki}, {Dufoer}, {Mizutani}, {Horiuchi}, {Tozuka}, {Takayama}, {Ohshima},
  {Saito}, {Dubovsky}, {Stone}, {Miller}, \& {Nogami}}]{2021PASJ...73.1209W}
{Wakamatsu}, Y., {Thorstensen}, J.~R., {Kojiguchi}, N., {et~al.} 2021, \pasj,
  73, 1209, \dodoi{10.1093/pasj/psab003}

\bibitem[{{Zamanov} {et~al.}(2004){Zamanov}, {Bode}, {Stanishev}, \&
  {Mart{\'\i}}}]{2004MNRAS.350.1477Z}
{Zamanov}, R., {Bode}, M.~F., {Stanishev}, V., \& {Mart{\'\i}}, J. 2004,
  \mnras, 350, 1477, \dodoi{10.1111/j.1365-2966.2004.07747.x}

\bibitem[{{Zamanov} {et~al.}(2016){Zamanov}, {Semkov}, {Stoyanov}, \&
  {Tomov}}]{2016ATel.8675....1Z}
{Zamanov}, R., {Semkov}, E., {Stoyanov}, K., \& {Tomov}, T. 2016, The
  Astronomer's Telegram, 8675, 1

\bibitem[{{Zamanov} \& {Zamanova}(1997)}]{1997IBVS.4461....1Z}
{Zamanov}, R.~K., \& {Zamanova}, V.~I. 1997, Information Bulletin on Variable
  Stars, 4461, 1

\end{thebibliography}
\bibliographystyle{aasjournal}

\end{document}